\documentclass{emulateapj} 

\begin{document}

\title{The Origin of Complex Behavior of Linearly Polarized Components in Parsec-Scale Jets}
\author{Philip A. Hughes}
\affil{Astronomy Department, University of Michigan, Ann Arbor, MI 48109-1090}
\email{phughes@umich.edu}

\begin{abstract}
Evidence that the magnetic fields of extragalactic jets have a significant
fraction of their energy in a random component is briefly summarized, and a
detailed model of evolving, jet polarization structures is constructed, based
on this picture.  The evolving magnetic field structure of an oblique shock
complex that forms in a relativistic jet simulation is explored by using
velocity data from the hydrodynamical simulation to advect an initially
random magnetic field distribution.  Radiative transfer calculations reveal
that emission from a propagating region of magnetic field, `ordered' by the
shock, and lying approximately transverse to the flow direction, merges with
that from an evolving sheared region at the flow periphery.  If such a flow
were barely resolved, observation would suggest evolution from a somewhat
oblique, to a more longitudinal, magnetic field structure with respect to the
flow axis, while higher resolution observations would infer a component
following a non-linear trajectory, and with a magnetic field orientation that
rotates during evolution. This result highlights the ambiguity in
interpreting VLBP data, and illustrates the importance of simulations in
providing a framework for proper interpretation of such data.
\end{abstract}


\section{Introduction}

 Relativistic jets are known to be a ubiquitous feature of environments where
accretion onto a compact object leads to outflowing material: they are
observed directly in Galactic sources, e.g., \cite{fen}, and in active
galaxies, e.g., \citet{zen}; and their presence is inferred for gamma-ray
bursts \citep{dar}.  While extraordinarily high temporal resolution of
activity in Galactic sources has been achieved using observatories such as
RXTE, e.g., \citet{bel}, the parsec-scale jets of AGNs remain crucial for the
exploration of evolving, spatially-resolved structures \citep{kel}.
Furthermore, linear polarization observations, e.g., \citet{ojh}, of such
sources provide valuable information about flow magnetic field structures,
and thus indirectly about the flow dynamics. The interpretation of these data
is complicated, because allowance must be made for line-of-sight integration
through a partially opaque source and relativistic effects; furthermore, it
is not yet clear whether magnetic relaxation, shear or oblique shocks are
primarily responsible for the observed magnetic field order, and
observational signatures capable of distinguishing these processes are yet to
be thoroughly quantified \citep{all}.

 The low (few percent) degree of polarization exhibited by compact
extragalactic radio sources when in a quiescent state, and the modest degree
of integrated polarization (rarely more than ten percent) during outbursts,
has been widely interpreted as due to `root-N' depolarization in a
synchrotron source with many randomly oriented magnetic cells within the
telescope beam. In particular, \citet{jon}, working with data between 1.4 and
90 GHz, at the upper end of which range opacity and Faraday effects should be
negligible, were able to demonstrate using Monte Carlo simulation that
observed `rotator events' arise naturally as a consequence of a random walk
as turbulent field is advected across observable part of the flow channel.
In developments of this picture, \citet{hua,hub,huc}, and \citet{gom,go2},
and \citet{go3} have successfully modeled the temporal, spatial, and spectral
attributes of outbursts in a number of individual sources, with a scenario in
which the shock compression of a flow provides an effective order to the
otherwise random magnetic field, and thus increases not only the total and
polarized fluxes, but also the percentage polarization. For the best-studied
source, BL~Lac, the model is particularly convincing, because the same flow
speed, orientation and compression are needed to independently reproduce the
outburst strength, temporal profile, and percentage polarization. While it
has proven more difficult to model other outbursts [even more recent
outbursts in the same source \citep{ala}; it appears that the earliest work
in this field was fortuitously modeling transverse shock events, while many
-- perhaps the majority of events -- involve oblique shocks \citep{all}],
accounting for the shock orientation has yielded promising results: as shown
by \citet{all}, even highly oblique shock structures may yield the observed
level of polarized emission due to shock compression, when allowance is made
for relativistic aberration. This picture has been strengthened by the work
of \citet{tom}, who argued that the absence of large circular polarization
severely limits the distribution of low energy particles; thus even if the
jet is an electron-ion rather than pair plasma, there is a severe limit to
the ability of Faraday effects to produce low levels of linear polarization,
which is more likely to result from `root-N' depolarization.  More recent
work allows for some of the magnetic energy to be in an ordered component of
magnetic field \citep{bec,hom}, but still requires a significant random
component.  The first results from the MOJAVE project \citep{lis} find
enhanced magnetic field ordering at flow bends, where shear would be expected
to be most influential in modifying an otherwise disordered magnetic field.
And, unless these sources are substantially beamed, Chandra and ASCA X-ray
observations of kiloparsec scale jet-knots imply magnetic fields as much as
four orders of magnitude in energy density below equipartition value
\citep{kas}. This suggests a dynamically unimportant magnetic field, subject
to the dynamical influence of entrainment and the onset of turbulence. 
It would be surprising if the character of the flow on a scale of 10-100s
parsecs is radically different.

Contrary to recent claims \citep{lyu} made in support of evidence for ordered
parsec-scale jet fields, a continuous sequence of shocks is not needed to
yield non-negligible interknot polarization.  Indeed, hydrodynamic
simulations, with the most minimal disturbance to the jet, reveal a complex
web of weak oblique structures that can impose a degree of order consistent
with the observed polarization: little compression is required in order to
put a fraction of the field energy in an effectively ordered form, and a
significant fraction of the field energy may be in a turbulent component,
while still admitting significant net polarization. The macroscopic
propagating knots are then distinct features imposed on this underlying
pattern, and their presence implies a spectrum of disturbances to the flow
that will simultaneously generate a degree of `background' order.  Thus,
while it is somewhat counterintuitive that in a flow subject to shear and
repeated shocking, magnetic field structure would remain highly disordered
three orders of magnitude (or more) in length scale beyond the `central
engine', multiple lines of evidence, accumulating since the early 1980s,
point to the fact that parsec-scale jet fields are highly disordered, and
that the discrete structures observed, form in such flows.  This
observational constraint has motivated the methodology applied in the work
discussed below, which assumes that the flow dynamics is not dominated by the
magnetic field, and that the magnetic field can therefore be evolved
independently of the hydrodynamics, in a `post-processing' step, once the
evolution of the hydrodynamic state variables has been established.

In principle, three dimensional relativistic magnetohydrodynamic (RMHD)
simulations of jets can make a major contribution to our understanding of
their evolution and propagation, and provide a basis for interpretation of
the data. However, rigorous quantification of the emergent radiation in all
Stokes parameters requires not only a detailed knowledge of the MHD
variables, but also of the radiating particle distribution; the only
simulations \citep{tre} to follow the nonthermal particles are not
relativistic.  Furthermore, while multidimensional RMHD simulations are being
performed, the emphasis is on understanding jet production by ergospheres and
disks, and considers only ordered, not turbulent magnetic field
\citep{vap,kob,koi}.

 Apart from the technical challenges of numerical RMHD, in particular
simultaneously preserving the divergence-free constraint on the field and
conserving its flux, there are two purely astrophysical issues, which are
particularly problematic for turbulent magnetic fields. First, interesting
structures (e.g., oblique shocks) may form in the simulated flow, but the
response of a turbulent magnetic field to the formation of such structures is
difficult to judge using conventional numerical RMHD methods, because an
initially turbulent field injected with jet material may have been subjected
to other compressions and shear (providing a significant degree of order,
which in a real flow might have been offset by entrainment or the spontaneous
generation of turbulence in a high Reynolds number medium), before the
structure of interest forms. Models for this may certainly be constructed,
but the details of the turbulence are poorly known, and deconvolving the
effects of a history of compressions and shear is very difficult. Second,
unknown microphysics will determine the resistivity of the flow, and thus the
rate of field line reconnection and relaxation. Over or under-estimating the
resistivity for a numerical simulation will lead to potentially spurious
field morphology far from the point of injection, also providing a
significant degree of order prior to formation of interesting structures.
The point of this study is to investigate the polarization structures that
result from the formation of particular flow structures (e.g., oblique
shocks), in the context of an initially random magnetic field.  To avoid the
aforementioned problems, epochs have been identified in the jet simulations,
that span the formation and propagation of distinct structural features (such
as oblique shocks), made spatial and temporal cuts to isolate such evolving
features, and used the corresponding velocity information to passively evolve
initially turbulent field distributions. This allows us to model the magnetic
structures that result from the response of an initially turbulent field to
the development of local hydrodynamic features -- the scenario strongly
supported by observations, as discussed above.

\section{Hydrodynamic Simulations}

 The results presented below were obtained using the fully 3D, relativistic,
hydrodynamic code for which the numerical method (RHLLE), tests, and first
results were discussed in detail by \citet{hmd}. The only two modifications
for the current application are as follows.  First, the code has been adapted
to run on a Linux cluster using CactusCode (http://www.cactuscode.org/); this
has required minimal code modification (the addition of {\it includes} and
{\it declarations} to the Fortran 90 source, the declaration of global
variables as the appropriate Cactus types, and the construction of the
appropriate {\it interface}, {\it parameter} and {\it schedule} units).
Shock-tube tests have been used to verify the integrity of the solver in this
new environment. The run described here was performed on $16$ processors of a
cluster. Second, in anticipation of modeling jets propagating into cluster
atmospheres, `pseudo-gravity' terms have been added to the solver, in order to
maintain an initial pressure gradient in the ambient medium.  I assume an
inviscid and compressible gas, and an ideal equation of state with constant
adiabatic index, $\Gamma$, and evolve mass density $R$, the three components
of the momentum density $M_x$, $M_y$ and $M_z$, and the total energy density
$E$ relative to the laboratory frame.  Defining the vector (in terms of its
transpose for compactness)
\begin{equation}
U = (R,M_x,M_y,M_z,E)^{T} ,
\end{equation}
and the three flux vectors
\begin{equation}
F^x = (R v^x,M_x v^x + p, M_y v^y, M_z v^z, (E + p)v^x) ^{T } ,
\end{equation}
\begin{equation}
F^y = (R v^y,M_x v^x, M_y v^y + p, M_z v^z, (E + p)v^y) ^{T } ,
\end{equation}
\begin{equation}
F^z = (R v^z,M_x v^x, M_y v^y, M_z v^z + p,(E + p)v^z) ^{T } ,
\end{equation}
where the three components of velocity are $v^x$, $v^y$ and $v^z$, the
conservative form of the relativistic Euler equation is
\begin{equation}
{\partial{ U}\over \partial{t}} + {\partial\over \partial{x} } (F^x) +
{\partial\over \partial y} (F^{y})+ {\partial\over \partial z} (F^{z}) = 0 .
\end{equation}
(The pressure, $p$, is given by the ideal gas equation of state in terms of
the internal energy, $e$, and mass density, $n$, by $ p = (\Gamma - 1)
(e - n) , $ and $c=1.$) To model a gravitational field that maintains an
arbitrarily imposed pressure gradient, I add a source term of form
$-\left(E+p\right)g$ to the $M_z$ update, and a source term of form $-M_zg,$
where $g=-\nabla p/\left(e+p\right)$, to the $E$ update. The effectiveness of
this approach was easily tested, by noting the absence of evolution of an
initial pressure gradient, in the absence of jet inflow, for a number of
computational cycles comparable to that used in the final simulation.

 The simulation discussed here is a low resolution trial set up to explore
the energization of cooling cores by AGN jets, and spans 4480 computational
cycles on a grid of extent $240\times 240\times 500$ cells; the lateral
extent of the domain is $\sim 10$ jet-radii, and the longitudinal extent is
$\sim 21$ jet-radii.  Both jet and ambient medium are characterized by an
adiabatic index $\Gamma=4/3$, the inflowing jet has (modest) Lorentz factor,
$\gamma=1.5$ and relativistic Mach number ${\cal M}=4.3$, and an opening
angle of $2\deg$. A precessional perturbation of lateral velocity $0.05$
and frequency $0.15$ (in units set by inflow speed and jet radius) is
applied to stimulate the development of normal modes of the cyclindrical
flow. The ambient medium has a density at the inflow plane $10\times$ that of
the jet, and declines with ambient pressure, corresponding to an isothermal
atmosphere. Pressure in the ambient medium is initially equal to that of the
inflowing jet material, and declines along the jet direction as
\begin{equation}
p\left(z\right)=p_0\left(\frac{z_c}{z/R_j+z_c}\right)^{\beta},
\end{equation}
where $R_j$ is the jet radius, $z_c$ is a scale height, taken here to be
$5R_j$, and the power law index was taken to be $0.6$ (Eilek, private 
communication).

\begin{figure}
\epsscale{0.65}
\plotone{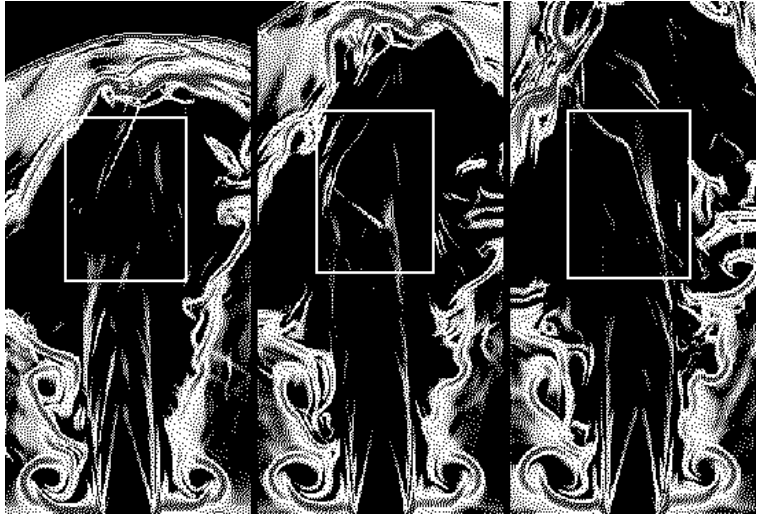}
\caption{
Schlieren renders (the gradient on an exponential map) of the laboratory
frame mass density for the simulation described in the text. The three panels
from left to right span the formation and evolution of an oblique shock
within the flow. The region extracted for study of magnetic field evolution
is shown by the superposed rectangles.
\label{fig1}}
\end{figure}

 Figure~\ref{fig1} shows the formation and evolution of an oblique shock in
the body of the jet, from computational cycle 17520 to 22000. The physical
time spanned by this stage in the flow's evolution is $9.8$ jet-diameter
light crossing times, and the pressure and density jumps along a normal to
the shock plane (evident in the figure) at its central point are respectively
$1.77$ and $1.46$.  Following \citet{mio}, it is assumed that the radiating
particle density varies as thermal pressure, and assuming a jump in the
magnetic field corresponding to the density jump, the shocked flow would have
emissivity enhanced by $\sim 3.1$ (assuming an optically thin spectral index
$\alpha=0.5$ ($S\propto \nu^{-\alpha}$), and ignoring small relativistic
effects), making this a good candidate event to model a parsec-scale jet
component. (This particular compression feature moves marginally subluminally
-- $\beta_{\rm max}\sim 0.3$ -- but its speed has no impact on the results
and conclusions presented below.) It is, therefore, interesting to explore
how a magnetic field, that is random at the start of the period studied,
would respond to the formation and evolution of this structure.  The
rectangles in Figure~\ref{fig1} outline the region selected for exploration
of magnetic field evolution. As the hydrodynamic data are extracted for use
in a post-processing step, and it is necessary to retain sufficient time
resolution, to keep the computation tractable, the velocity data are averaged
over cubes of size $4^3$ cells, and saved $225$ time slices, each of size
$30\times 30\times 40$ cells.

\section{Magnetic Field Initialization and Evolution}

 An initially random magnetic field was generated by selecting random phases,
and random amplitudes from a Rayleigh distribution, for the Fourier transform
of the vector potential, $\tilde{\bf A}\left({\bf k}\right)$. The Fourier
transform of the magnetic field is then $\tilde{\bf B}\left({\bf k}\right)
=i{\bf k}\times \tilde{\bf A}\left({\bf k}\right),$ and an inverse Fourier
transform of this yields a magnetic field that is guaranteed to satisfy the
divergence-free constraint. (See, e.g., \citet{tri} for a detailed exposition
of this approach.) By choosing a Gaussian form for the variance, $\sigma^2
\left(k\right),$ in the Rayleigh distribution, the correlation scale can be
controlled, and it was chosen to be $4$ computational cells -- 10\% of the
computational volume height; `root-N' depolarization will then lead to a net
polarization for the total emission from the volume of order a few percent,
typical of the quiescent state of radio-loud AGN \citep{jon}. 

 Over the $\sim 3$ light crossing times covered by the evolution followed
here, magnetic field is advected primarily along the jet axis -- i.e., in the
$z$ direction. Thus, to provide random field to be advected into the `active'
computational volume from below, a $30\times 30\times 120$ field distribution
(cut from a $120^3$ data cube to ensure that asymmetry did not induce power
preferentially along any one axis) was generated; velocity data from the
hydrodynamic simulation were associated with the uppermost $40$ planes of
cells. At each time step, the lowest-$z$ velocity data were copied into what
is effectively an extended boundary of $80$ planes of cells below the inflow
plane of the hydrodynamic simulation; velocity and magnetic field components
were copied into a single-cell-wide ghostzone along each of the other faces
of the volume.

 The magnetic field is advected according to the induction equation,
\begin{equation}
{\partial {\cal B}\over \partial t}=\nabla\times \left({\bf v}\times {\cal B}
\right)+\eta \nabla^2{\cal B},
\end{equation}
where $\eta$ is the resistivity (the dissipative term being added for
computational expediency -- see below -- by analogy with the nonrelativistic
induction equation), and the field ${\cal B}$ relates to the spatial
components of the magnetic four vector, ${\bf B}$, through
\begin{equation}
{\bf B}={{\cal B}\over\gamma}+\gamma\left({\cal B}\cdot{\bf v}\right){\bf v},
\end{equation}
for Lorentz factor $\gamma.$ For numerical purposes the induction equation is
recast in integral form:
\begin{equation}
{\partial \over \partial t}\int_{\Sigma}{\cal B}\cdot d{\bf S} +
   \oint_C \left({\cal B}\times {\bf v}+\eta \nabla \times {\cal B}\right)\cdot
   d\,{\bf l}=0.
\end{equation}
Discretization then yields algebraic equations for the updates to the average
of the magnetic field over a surface, proportional to the discretization
scales as $\Delta t/\Delta h.$ The velocities are regarded as stored at cell
centers, and the magnetic field components on cell faces -- the surface for
integration (the $x$-component on the higher $x$ face of a voxel, etc.). The
line integral thus becomes the sum of terms around face edges, and the
velocity and magnetic field component values at the centers of these edges
are derived by interpolation as needed.  Upwinding is not relevant here, as
the velocity data are already known; fluxes are not computed from the
characteristic speeds. (See, e.g., \citet{koa} for a detailed exposition of
this approach.) This technique preserves the divergence-free constraint to
machine accuracy, but unfortunately does not conserve flux. It has been
established by advecting a random field with a {\it uniform} velocity
distribution, that by adding a small dissipation, $\eta\sim 0.085,$  the
global magnetic energy density does not grow, while the field remains random
(as judged by the variance of the components) to within 15\% at the mid-time
in evolution, when structure of interest develops.

  Figure~\ref{fig2} shows the fluid frame magnetic field components in a
$x={\rm const}$ plane (the same midplane as that shown in Figure~\ref{fig1})
at $9$ epochs during evolution by the pre-computed velocity field. In each
panel the length of the lines is a measure of the average magnetic field in a
$3^3$ cube centered on the cut plane, and their maximum length is the same in
each panel, so that display of the evolution of field amplitude is suppressed
to emphasize changes in the field direction. The orientation of the lines
shows the overall sense of the field within each local $3^3$ subvolume; this
is computed from $\tan^{-1}\left(B_z^T/B_y^T\right),$ where $B_y^T=\Sigma
B_i\cos\left(2\phi_i\right),$ and $B_z^T=\Sigma B_i\sin\left(2\phi_i\right),$
the angles $\phi_i$ being the local field orientation to the $y$-axis. 

  This figure does not therefore show the degree of {\it order} of the
magnetic field (a field that is almost random locally will, in general,
display some net sense, and can be of arbitrary strength), which will
determine the degree of polarization of emergent radiation.  The degree of
polarization measured by an observer is not a local property of the field;
within each $3^3$ subvolume, the adopted correlation imposes a degree of
order, and the polarization of the emergent radiation will depend upon
integration along some chosen line-of-sight through a number of planes such
as shown here. A measure of the field order has, however, been computed by
looking at the rms values of the field components parallel and perpendicular
to the preferred sense of the field in each $3^3$ subvolume, and it is found
that the field order is a complex function of position. For example, shear
imposes a significant degree of order on the field even in the outlying
regions of much lower magnetic field strength, implying that a measure of
field order alone cannot be used to estimate the regions from which a high
flux of polarized radiation results, since the intrinsic emissivity of these
weaker field regions will be low. Our measure of field order does confirm
that the field is ordered in the downstream flow of the oblique shock, by an
amount that would be expected from the density jump, which is the main
feature of interest here.  The appearance of these structures will depend on
their relative Doppler boosting; however, the interest here is to model knots
that display measurable motion, in which case we will not be seeing the flow
far from its critical cone -- and statistically be more likely to be outside
of that cone -- so boosting may be somewhat limited.  Thus, although a full
radiative transfer calculation must be performed to understand in detail how
these simulations relate to VLBP maps, the extended regions of well-defined
field structure seen in the later panels of Figure~\ref{fig2} can be expected
to correspond to knots of high intensity and polarization.

\section{Radiative Transfer}

 The transfer of polarized radiation through a diffuse plasma, allowing for
emission, absorption, the birefringence effects of Faraday rotation and
mode conversion (which can produce modest levels of circular polarization),
and relativistic aberration and boost, has been described in detail by a
number of authors. I follow the analysis of \citet{jod}, which is compactly
summarized by \citet{tom}, and which has been used previously in the study of
parsec-scale jets; see, for example, \citet{hua} and \citet{gom}.

 An algorithm for radiative transfer has been constructed as follows:  the
observer may be placed at arbitrary polar angles ($\theta$,$\phi$),
defined in the conventional sense with respect to the Cartesian system used
to describe the hydrodynamics; thus an observer viewing the $x={\rm const}$
plane of Figure~\ref{fig2} will be at polar coordinate ($90$,$0$). An array
of lines-of-sight is established, using a preset density of lines along the
longest axis of the projection of the computational volume on the plane of
the sky, with a commensurate number orthogonal to that, to ensure equal
resolution in the two directions. For the results presented below, a
resolution of $128$ pixels along the longest axis was used. For each
line-of-sight, the algorithm finds the most distant cell, and radiative
transfer is performed, cell to cell, until the `near side' of the volume is
exited. Within each cell, an aberrated magnetic field direction is computed
from the rest frame field, velocity, and observer location, and is used with
transfer coefficients modified by the relevant Doppler boost, as described by
\citet{tom}.

 A full radiative transfer requires some knowledge of the particle
distribution function: the low energy cutoff Lorentz factor, the Lorentz
factor of particles radiating at the observer frequency, and the spectral
slope of the distribution; these are arbitrarily fixed at $\gamma_{i0}=100$,
$\gamma_{n0}=1000$, and $\alpha=0.75$ (see \S 2).  In principle these might
change from place to place, but as no model of this is available, these are
regarded as fixed parameters, and when exploring optically and Faraday thick
flows, the predicted behavior as illustrative only.

 The transfer coefficients are computed by assuming that the radiating
particle density is proportional to the thermal pressure of the hydrodynamic
simulation; however, one cannot a priori know the optical depth through some
arbitrary line-of-sight.  The magnetic field strength and hydrodynamic
pressure are scaled by their rms value and mean value respectively (within
the computational volume at the initial time), and the optical depth is
adjusted by scaling to yield some target optical depth for a line-of-sight
with the rms field and mean pressure through the longest extent of the volume.
For the results presented below, values were chosen to yield an optically
thin flow.

 Observations strongly suggest that the particle distribution and magnetic
field, jointly responsible for the synchrotron radiation, are confined to the
inflowing jet, and do not permeate the ambient medium. To model this,
a filter is applied, so that no emission arises from cells with velocity
$v<v_{\rm crit}=0.5$ ($c=1$). The intensity pattern on the plane of the sky
has been convolved with a $10$-pixel Gaussian beam -- corresponding to four
beam widths across the shock structure of interest, and typical of the best
VLBI resolutions now being attained for parsec scale jets, where in some
sources it is now possible to detect limb brightening, and rotation measure
gradients, across the flow -- and all four Stokes parameters ($I$, $Q$, $U$
and $V$) stored in FITS files, for subsequent display using Dan Homan's
pgperl package FITSplot (http://personal.denison.edu/$\sim$homand/\#useful).
In principle, the radiative transfer should use retarded times; that has not
been done in the results shown below, but this will have minimal impact on
the validity of the results, as individual features move rather slowly.
Rigorous, retarded time calculations, will be performed in a subsequent
study.

Figure~\ref{fig3} shows, from upper-left to lower-right, `fake' maps of
polarized intensity, corresponding to the $5$th through $8$th epochs shown in
Figure~\ref{fig2}, computed for an observer orientation of ($90$,$20$), with
other parameters as described above; this corresponds approximately to the
view through the midplane of the computational volume shown in
Figure~\ref{fig2}.  Care must be taken in comparing Figure~\ref{fig2} with
Figure~\ref{fig3}:  the former is a slice through a computational volume,
whereas the latter shows an integration along lines of sight; the `fake' maps
are made for a viewer not precisely along a coordinate direction, slightly
changing the aspect ratio of the panels; in general, retarded time
calculations are needed, and there is no simple connection between the
sequence of frames in the two figures. As retarded time calculations have not
been employed here, the corresponding frames in the two sequences have been
labeled accordingly.  The scaling has been chosen so that the maps have
comparable dynamic range. Corresponding maps of total intensity have the same
general morphology, and I choose to discuss the $P$ map because the
components are slightly more distinct. Corresponding polarization maps for a
flow exhibiting significant opacity have quite different structure, but in
that case the percentage polarization is negligible; the structure results
from lines-of-sight with, by chance, some residual polarization, and the
pattern bears no simple relation to the underlying MHD. (This may provide a
useful model for the exploration of low polarization cores, but here we are
interested in exploring the evolving structures evident on maps in regions of
low optical depth.) Maps made with different observer orientation can look
radically different, because a few spatially distinct peaks of emission
within the 3D volume are seen in quite a different projection. That varied
and complex behavior can result from simple flows is one of the messages of
this work; I opt to present the map sequence that can be most easily related
to the underlying flow features.

 The localized peak in intensity in the first panel of Figure~\ref{fig3} is
offset from the regions of most-ordered field at that epoch -- associated
with the shock above, and shear at right -- seen in the central panel of
Figure~\ref{fig2}, because that line-of-sight intersects regions of ordered
field and high pressure not revealed by the simple center-plane cut of
Figure~\ref{fig2}.  Careful exploration of the flow structure shows that what
appears to be a simple, oblique shock, does indeed comprise thin,
approximately planar features oblique to the flow direction, but that there
are two nested discontinuities, which curve slightly to follow the local jet
boundary, extending for about one-third of the jet circumference, and
converging at their base on the flow edge. The devil is in the detail, and in
this case, the devil is in the knot of emission. Even the simplest of VLBI
structures may be hiding a complex flow pattern. Subsequent evolution, shown
in the upper-right, lower-left, and lower-right panels of Figure~\ref{fig3},
leads to a weak feature downstream of the initial bright knot, reflecting the
increasing influence of emission from both the downstream flow of the shock
`complex', and the shear at the jet edge above this, and, finally, another
distinct knot associated with the region of strong shear -- evident in the
upper left of the last panels of Figure~\ref{fig2} -- that arises as the jet
direction changes. That effect propagates in the flow's upstream direction,
elongating this knot. 

 This simulation is applicable to the interpretation of sources such as
BL~Lac objects, where components are short-lived, and where, within a few
beam widths -- which barely start to resolve these components, leading to
maps with structure smoother than one might anticipate -- complex dynamical
behavior can be seen, with variable apparent speed and/or nonlinear
trajectories \citep{den}.  No claim is made that this simulation explains the
behavior of any particular VLBI source, but it is interesting to see that an
asymmetric shear layer is easily produced, which might have relevance for the
interpretation of VLBI features suggestive of the presence of a flow boundary
\citep{atr}, and it clearly warns against the naive interpretation of
observations that reveal knots that seemingly abruptly change direction and
decelerate. Fortunately, the sense of the polarized emission would provide
some help in disentangling the behavior displayed here, if these were actual
data: the electric vector of the knot of emission in the first panel of
Figure~\ref{fig3} is highly inclined to the average flow direction, being
orthogonal to the oblique structure shown in the corresponding schlieren
render of Figure~\ref{fig1}, whereas the electric vector for the emission
evident in the last panel is orthogonal to the flow boundary. This gives
strong clues as to the magnetic field morphology (at least for an optically
thin, weakly relativistic, flow) and thus to the underlying hydrodynamics.

\section{Discussion}

 The structure and dynamical significance of extragalactic jet magnetic
fields has received much attention over the last years, with considerable
emphasis on the case of fields with large scale order, and a tendency to
dismiss the possibility of a significant random component. Lessons from
studying the magnetic fields of other astrophysical bodies suggest that such
an `either-or' view is misplaced, and that it is most likely that there is
magnetic energy on a range of scales: both ordered and random components may
play a role in determining the emission properties of many sources, or one
may dominate under certain conditions. Strong evidence exists for the
prescence of a random component, as discussed in the introduction, and we
have explored some of the observational consequences of that picture.  The
simulation discussed above was chosen as a reasonable data set with which to
develop the technology for field evolution and radiative transfer.  Results
should be regarded as illustrative, and not over-interpreted. The flow is of
low Lorentz factor and low Mach number, the shock that forms is weak, and the
resultant field compression produces very modest levels of polarization.
However, it does clearly expose the consequences of having spatially
distinct, varying centers of activity, yielding a complex behavior, the
character of which is very sensitive to the observer's orientation.  The
hydrodynamic structure will be similar to that seen here, for Lorentz factors
up to about 5 \citep{dun,ros}, while faster flows will in general be rather
featureless. At speeds higher than that addressed here, a major role will be
played by relative Doppler boosting, which can be expected to lead to a
detailed pattern of intensity that is very dependent on observer orientation;
this will probably lead to an even more complex interplay between regions of
compression and shear, but only detailed modeling for a range of parameters
can quantify this.

\begin{figure}
\epsscale{0.65}
\plotone{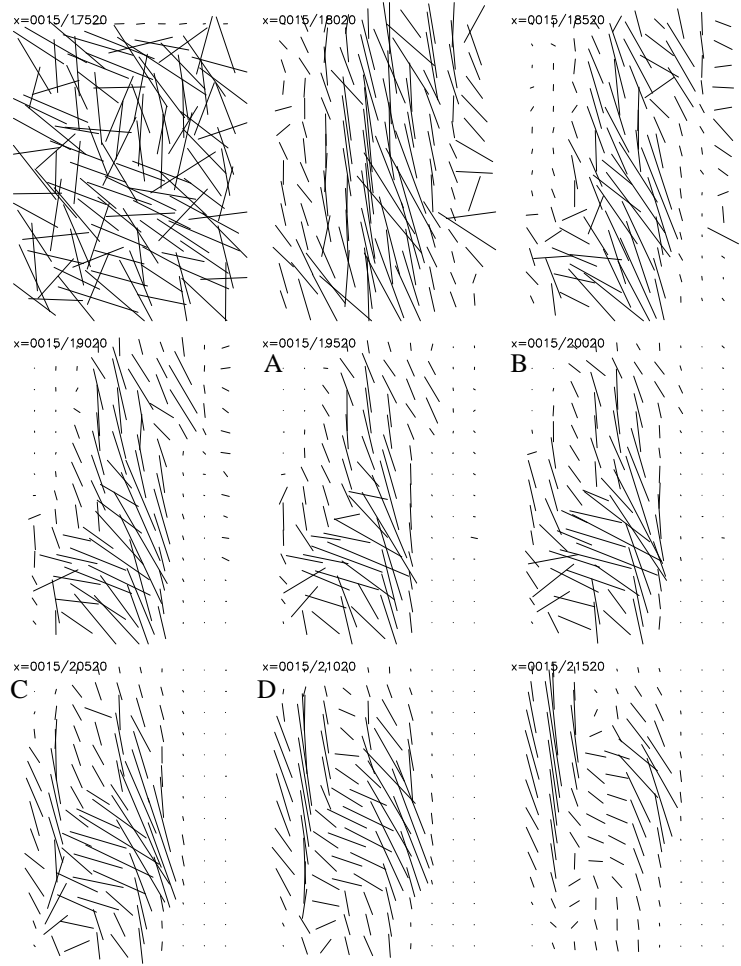}
\caption{
From left to right and top to bottom, the panels show the magnetic field
during $9$ epochs of its evolution during advection by a pre-computed
velocity field. The orientation of the lines gives the sense of the dominant
magnetic field within regions of $3^3$ cells on the same midplane as shown in
Figure~\ref{fig1}. The length of each line indicates the local magnetic field
strength. The central panel corresponds to the central panel of
Figure~\ref{fig1}. The panels labeled `A', `B', etc. correspond to the
panels of Figure~\ref{fig3}.
\label{fig2}}
\end{figure}

\begin{figure}
\epsscale{0.65}
\plotone{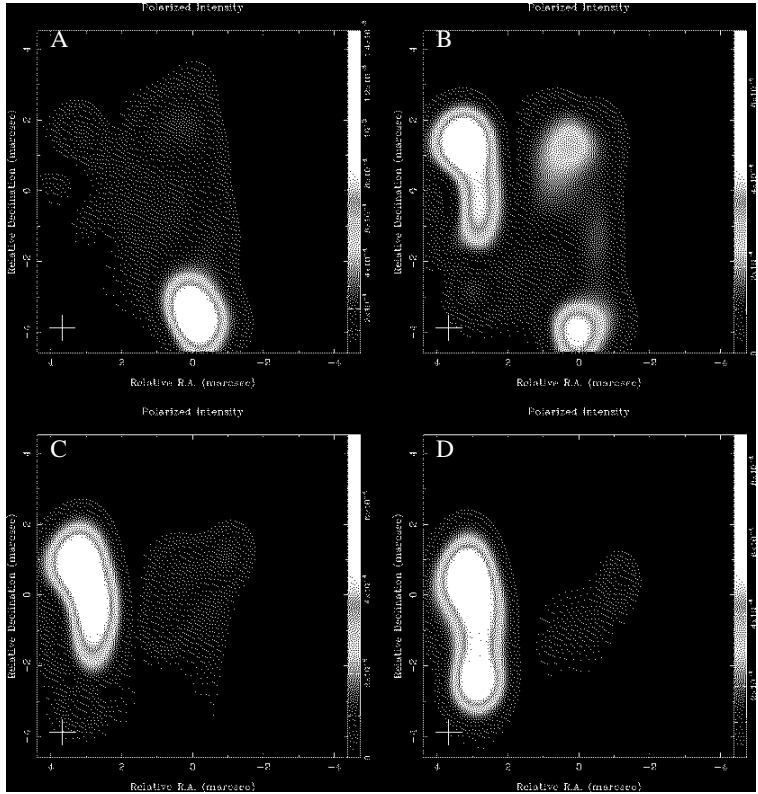}
\caption{
From upper-left to lower-right, `fake' maps of polarized intensity, computed
by performing radiative transfer calculations through the full computational
volume for which the $5$th through $8$th epochs shown in Figure~\ref{fig2}
(labeled `A', `B', etc.) are slices.  Features on these maps may be related
to features seen in Figure~\ref{fig2}, for example, the ridge of emission in
the lower-right panel corresponds to the sheared region of longitudinal field
seen to the left of the flow in the last panel of Figure~\ref{fig2}. Major
differences between the two figures arise because the radiative transfer is
probing flow not revealed by the simple cut of Figure~\ref{fig2}.
\label{fig3}}
\end{figure}

 Examination of Figure~\ref{fig2} shows that a region of ordered, obliquely
oriented, magnetic field arises in the downstream flow of the oblique shock
which is well-defined by the mid-time in evolution. This structure propagates
subluminally along the flow direction, weakening as it evolves, and, most
strikingly, merges with the more oblique, wedge-shaped region dominated by
shear, most obvious above center-right in the last panel. Radiative transfer
reveals (as shown in Figure~\ref{fig3}) an evolving pattern of emission (both
total and polarized) that has somewhat different morphology from that which
might be inferred from the sampling of the magnetic field structure, because
lines-of-sight probe flow structure hidden by simple data slicing, and
incorporate the effects of radiating particle density and line-of-sight
depth. However, whether based on MHD structures alone, or accounting for a
full radiative transfer, the conclusion is that the interrelated evolution of
internal jet structures and shear layers can lead to apparently complex
behavior in which the changing relative importance of these facets of the
flow mimic the evolution of a single component with complex dynamical
behavior.  It is important to realize that the domains dominated by shear at
the jet edge are not static, and that these regions of shear-influenced field
evolve as the jet shifts laterally. If the flow cross-section were barely
resolved by observation, it would appear that a region of more-nearly
transverse field evolved into a region of longitudinal field; higher
resolution might suggest that a component within the flow propagates along a
somewhat curved trajectory, with a rotation of the plane of polarized
emission towards the longitudinal sense.

 Simulations such as presented here caution against the naive interpretation
of multi-epoch VLBI/P maps -- in particular showing that non-linear
trajectories and a rotation of the plane of polarization as a component
evolves may be manifestations of evolution {\it within} a channel broader
than that suggested by the data -- and the simulations provide a framework
within which to interpret them. The global flow pattern may be more
complicated than the trajectory of one knot suggests. The simulations also
serve to clarify that evolving, complex, internal structure, that may be
readily identified on a time series of VLBP maps, does not necessarily result
from the interaction of the jet with ambient density or pressure structures,
but, rather, may form and evolve due to the response of the jet to
infinitesimal perturbations.  Defining the dynamics of such flows through
VLBI observations clearly demands a sequence of maps with high time
resolution, and of sufficient duration to probe the global structure of the
flow. Isolated epochs may reveal transient peculiarities, not representative
of the overall flow morphology.

\acknowledgments

This work was supported in part by NSF grant AST 0205105.  The work was
motivated in part by results from the UMRAO monitoring program supported in
part by a series of grants from the NSF and by funding from the Department of
Astronomy, and the author thanks Hugh and Margo Aller for illuminating
discussions on the observational implications of the study. Comer Duncan and
Mark Miller have made major contributions to the hydrodynamic simulations.
A referee made numerous comments that helped to significantly improve the
manuscript.

\clearpage

{\bf This figure is a low resolution placeholder for astro-ph. 
The original may be found at 
http://www.astro.lsa.umich.edu/users/hughes/icon\_dir/relproj.html\#MAG}

\clearpage

\clearpage

{\bf This figure is a low resolution placeholder for astro-ph. 
The original may be found at 
http://www.astro.lsa.umich.edu/users/hughes/icon\_dir/relproj.html\#MAG}

\end{document}